\documentclass[twocolumn,showpacs,amsmath,amssymb,pra]{revtex4}
\usepackage{epsfig}
\usepackage{color}
\usepackage{graphicx}
\usepackage{bm}
\usepackage{epstopdf}

\begin{document}

\title{Borromean ground state of fermions in two dimensions}

\author{A.~G. Volosniev} 
\author{D.~V. Fedorov} 
\author{A.~S. Jensen}
\author{N.~T. Zinner}
\affiliation{Department of Physics and Astronomy, Aarhus University, 
DK-8000 Aarhus C, Denmark} 

\date{\today}

\begin{abstract}
The study of quantum mechanical bound states is as old as 
quantum theory itself. Yet, it took many years to realize that 
three-body borromean systems that are bound when
any two-body subsystem is unbound are abundant in nature.
Here we demonstrate the existence of borromean
systems of spin-polarized (spinless) identical fermions in two spatial
dimensions.  The ground state with zero orbital (planar) angular
momentum exists in a borromean window between critical two- and
three-body strengths.  The doubly degenerate first excited states of
angular momentum one appears only very close to the two-body
threshold.  They are the lowest in a possible sequence of so-called
super-Efimov states.  While the observation of the super-Efimov
scaling could be very difficult, the borromean ground state should be
observable in cold atomic gases and could be the basis for producing a
quantum gas of three-body states in two dimensions.
\end{abstract}

\maketitle

Unlike classical mechanics quantum mechanics allows bound $N$-body
states without having bound subsystems.  These so-called borromean
systems are discussed in a number of publications for the simplest
example of three particles
\cite{efi70,zhukov1993,richard1994,richard2000,nielsen2001,jen04,bra06}.
The phenomenon was recently even generalized to more particles and
higher orders, see f.ex. Ref.~\cite{baas12}. 
Borromean three-body
systems are abundant in three dimensions (3D) for both bosonic and
fermionic systems in nuclear, atomic and molecular physics
\cite{macek2006,zinnernucl}.
However, the behavior is strikingly different for bosonic systems in one or
two (2D) spatial dimensions~\cite{nielsen1999,nielsen1997,volbor}.
where bound states appear for infinitesimally small attractions.
Without an artificial repulsive barrier at large distance it is
virtually impossible to form a borromean bosonic 2D system
\cite{nielsen1999,nielsen1997,volbor}.  Furthermore, the celebrated Efimov
effect of infinitely many bound three-body states at the two-body
threshold is not present in 2D \cite{bruch1979, vugalter1983}.

Three identical spin-polarized fermions are harder to bind than
bosons, because relative two-body $s$-states are forbidden by the
Pauli principle.  It would therefore intuitively be more difficult to
form borromean systems and the Efimov effect should be out of reach.
However, both effects are possible in 2D as we shall discuss in the
present paper.  The Efimov effect was recently derived in an extreme
double-exponential scaling form \cite{superefimov} using a momentum
space formalism.  The same scaling behavior was also found in
low-energy scattering of three spinless fermions \cite{levinsen}.  

In this paper we present analytical and
numerical evidence for a new state of fermionic matter where the basic
constituent is a borromean three-fermion cluster.  This may be
considered a trion quantum gas as opposed to the two-component BCS-BEC
crossover \cite{bloch2008,giorgini2008} driven by the absence (BCS) or
presence (BEC) of a two-body bound state.  The new structure should be
accessible through the tunability of both interactions and geometry of
modern cold atom experiments \cite{zinnernucl}. 
Both the single
particles and the three-body borromean states are fermionic entities
and thus have to obey the Pauli principle. This could imply an 
increase in stability of the many-body system, but estimates of the
lifetime show that it is expected to be 
considerably shorter than the two-component Fermi 
gas \cite{levinsen,pricoupenko2008}. The creation of a superfluid state of spin-polarized 
fermions is still an important outstanding goal in the field. 
Observation of the three-body borromean ground state is, however, 
not dependent on reaching the superfluid state. When formed, these
three-body states are likely susceptible to chemical reactions and
may be used to study quantum chemical dynamics of a fermionic 
gas with suppressed two-body but enhanced three-body reactions.

\section{Two fermions at the two-body threshold for binding}
Borromean states are most easily found for potentials that can almost
bind a two-body system.  The relative motion in two-body systems is
described with the wave function given as a product of radial,
$\phi_M(r)$, and angular, $\exp(iM\theta)$, parts, where $(r,\theta)$
are the polar relative coordinates, and $M=0,\pm 1, \pm 2,...$ is the
2D angular momentum quantum number.  The radial Schr\"{o}dinger
equation is
\begin{equation}
  \frac{\hbar^2}{m} \bigg(-\frac{1}{r}\frac{\partial}{\partial r}r\frac{\partial}{\partial r}+\
\frac{M^2}{r^2}\bigg)\phi_M  = (E_2-gV) \phi_M\; ,
 \label{shr-rad-eq}
\end{equation}
where the bounded potential, $V(r)$, of cylindrical symmetry is assumed to
decrease faster than $1/r^{2+\epsilon}$, $\epsilon>0$, for large $r$, 
or in practice treated as zero outside a finite radius $R_0$.  The two-body energy is
$E_2$, $g>0$, is a dimensionless strength parameter, and $m$ is the
mass of one particle.  The doubly degenerate antisymmetric ground
state for spin polarized fermions has $M=\pm 1$, since $M=0$ describes
a symmetric total wave function.  Therefore, for two fermions we shall
only consider $M=1$ and omit the related index.  The regular zero
energy solution, $\phi(r)$, to Eq.~\eqref{shr-rad-eq} obeys
\cite{newton1986}
\begin{equation}
\phi(r)=r-\frac{gm}{2\hbar^2}\int_0^r \mathrm{d}s
\frac{s^2-r^2}{r}V(s)\phi(s).
\label{funct-small}
\end{equation}
The large-distance asymptotic of this $E_2 = 0$ solution is uniquely
determined by the length parameter, $a$, defined such that $\phi(r)$
asymptotically approaches $(r^2-a^2)/r$ \cite{nielsen2001}, where
\begin{equation}\label{aeq}
a^2 = \dfrac{ \frac{gm}{\hbar^2}\int_0^\infty\mathrm{d}s s^2 
V(s)\phi(s)}{2+\frac{gm}{\hbar^2}\int_0^\infty \mathrm{d}s \phi(s)V(s)}.
\end{equation}   
The critical value, $g^{cr}_2$, for two-body binding is reached when
$a = \infty$ where the zero-energy solution crosses zero at
$r\to\infty$, see Methods.  Then $g^{cr}_2$ is defined as the
smallest value of $g$ where the denominator in Eq.~\eqref{aeq}
vanishes.  The limit $g\to 0$ corresponds to $a\to0$, since
binding of two identical fermions in 2D requires a finite attraction.

\subsection{Numerical results for two fermions}
To address weakly bound two-fermion systems we use
the stochastic variational method~\cite{suzuki1998,mitroy,volthesis}
to calculate ground state energies for different values of $g$. To 
illustrate the generic nature of our findings, we
choose three qualitatively different potentials
\begin{eqnarray}
 V_1(r) &=& -\frac{\hbar^2}{mb^2}e^{-r^2/b^2} \;, \nonumber \\
 V_2(r) &=& -\frac{\hbar^2}{mb^2}(e^{-r^2/b^2}-0.5e^{-0.5r^2/b^2})\;, \label{pot}\\
 V_3(r) &=& \frac{\hbar^2}{mb^2}(e^{-r^2/b^2}-0.8e^{-0.5r^2/b^2}) \; . \nonumber 
\end{eqnarray}

\begin{table}
\centering
\begin{tabular}{ | c | c | c| c| c | c| c| }
\hline
 $V$ & $g_2^{cr}$ & $g_3^{cr}/g_2^{cr}$ & $E_3(g_2^{cr})$ & $E_3^*(g_2^{cr})$ & 
$\langle \rho^2 \rangle_{gr}$ &  $\langle \rho^2 \rangle_{exc}$  \\
\hline
 $V_1(r) $ & $6.72$ & $0.72$ & $-1.50$ & $-0.18$ & $1.65$ & $6.0$ \\
 $V_2(r)$ & $28.98$ & $0.68$ & $-5.55$ & $-0.47$ & $0.56$ & $1.13$ \\
 $V_3(r)$ & $8.63$ & $0.72$ & $-0.439$ & $-0.045$ & $5.9$ & $22.7$ \\
\hline
\end{tabular}
\caption{Borromean binding energies.
	Numerically calculated two-body thresholds
  for binding, three-body properties at the thresholds and 
  estimates for three-body thresholds for binding. Ground and
  excited states have angular momentum $0$ and $\pm 1$. Lengths and
  energies are in units of $b$ and $\hbar^2/(mb^2)$, respectively. }
\label{tab}
\end{table}

The first has a simple attractive pocket, the second has a repulsive
barrier outside the attraction, and the third has a repulsive core.
The calculated values of $g_2^{cr}$ are given in Tab.~\ref{tab}.  The
two-body bound states with $E_2 \to 0$ are universal, independent of
potential, i.e. the wave function is a modified Bessel function
$K_1(|k|r)$ $( k^2=mE_2/\hbar^2)$, corresponding to $V=0$ with
essentially only non-zero probability in the classically forbidden
region of zero potential.  Using the wave function, $K_1$, in all
space we find the mean square radius for small $k$, i.e.
\begin{equation}
\langle r^2 \rangle |k|^{2}\ln(|k|b) = -\frac{2}{3} +O\left(\frac{1}{\ln(|k|b)}\right), 
\label{as-r}
\end{equation}
where we used the potential range parameter, $b$, to get a
dimensionless argument of the logarithm.  

\begin{figure}
\includegraphics[scale=1.0]{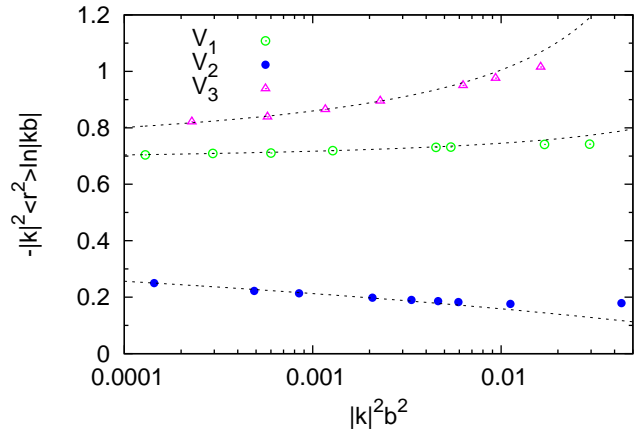}
\caption{Asymptotic behaviour of the two-body states.
	Calculated mean square radii as function of the two-body
  binding energy for the three potentials in eq.~(\ref{pot}). The
  points are obtained numerically by decreasing the attraction $g$ through the
  variation $g\to g_2^{cr}$. The curves are produced by fits with
  a length parameter, $b_0$, in eq.~(\ref{as-r}) different from $b$, i.e. we 
  have fitted the numerical results to the functional form on the left-hand side
  of eq.~(\ref{as-r}). This incorporates 
  next-to-leading order terms proportional to $\left[\ln(|k|b)\right]^{-1}$.
  The limit for $k=0$ is $2/3$. The parameters of the fits are $\ln(b_0/b)=0.4864$ 
  for $V_1$, $\ln(b_0/b)=1.55$ for $V_3$ and $\ln(b_0/b)=-14.74$ for $V_2$.}
\label{figrms}
\end{figure}

To confirm our understanding through Eq.~\eqref{as-r} is numerically
challenging for some potentials.  In Fig.~\ref{figrms} we see that the
limit of $2/3$ only is reached for very small $E_2$, although much
faster for $V_1$ and $V_3$ than for $V_2$.  The different rates of
convergence illustrate how small an energy is necessary to reach the
universality limit for potentials of varying structure.  The
similarity for $V_1$ and $V_3$ arises because the repulsive core for
$V_3$ is ineffective for the $p$-waves, while their attractions are
similar.  On the other hand, the external barrier for $V_2$ delays
tunneling into the universal region of the classically forbidden space
and beyond. The radius increases with decreasing $k$ as larger
distances are populated.  However, the universal limit will eventually
be reached for sufficiently small $E_2$.

These properties are markedly different for two bosons in $2D$ where
the symmetric wave function with $M=0$ is well described by
$K_0(|k|r)$ at comparably small binding energies \cite{vol11}.  These
observations are of crucial importance in understanding 
the universal two-body bound states that arise near the window for borromean 
binding.

\section{Three fermions}
The two-fermion binding requirement of a finite potential depth in
$2D$ resembles that of two bosons in $3D$.  This strongly suggests
that borromean three-fermion systems are possible in $2D$ even for
purely attractive potentials.  Also, the Efimov effect cannot
be strictly ruled out for three spin-polarized fermions in $2D$
\cite{vugalter1983}.  In contrast, we know that the Efimov effect is
absent for three bosons in 2D, and borromean states only occur for
potentials with positive net volume, $\int V(r)r\mathrm{d}r>0$, and
regions with substantial repulsion
\cite{nielsen1997,volbor,nielsen1999,bruch1979,vugalter1983}.

In 3D, the coordinate-space adiabatic hyperspherical expansion method
proved to be very efficient for short-range interactions due to small
large-distance couplings \cite{nielsen2001}.  This efficiency is
highlighted in the precise description of the Efimov effect by use of
only the lowest adiabatic potential
\cite{fedorov1993,jen96,macek2006}.  We shall therefore here employ
the 2D hyperspherical formalism
\cite{nielsen1997,nielsen1999,nielsen2001} for three spin-polarized
fermions.

Let ${\bf r_i}$ be the coordinate of the $i$th particle. One set of
the relative Jacobi coordinates is ${\bf x_i}=({\bf
  r_j-r_k})/\sqrt{2}$ and ${\bf y_i} = ({\bf
  r_j+r_k})/\sqrt{6}-\sqrt{2/3}{\bf r_i}$, while the other sets are
obtained by cyclic permutation of $\{i,j,k\}=\{1,2,3\}$.  The
hyperspherical coordinates are given by $\rho = \sqrt{{\bf x_i}^2+{\bf
    y_i}^2}$ and the three angles for each Jacobi set, $ \Omega_i =
\{\alpha_i, \Theta_{xi}, \Theta_{yi}\}$, where
$\alpha_i=\arctan{(x_i/y_i)}$, and $\Theta_{xi}$ and $\Theta_{yi}$
define directions of the coordinates $({\bf x_i},{\bf y_i})$.  The
kinetic energy operator has the form
\begin{align} \label{kinop} 
T=-\dfrac{\hbar^2}{2m}\left(\rho^{-3/2}\dfrac{\partial^2}{\partial \rho^2}\rho^{3/2} 
- \dfrac{3}{4\rho^2}\right) 
+ \dfrac{\hbar^2}{2m \rho^2} \Lambda^2 
\end{align}
where the hyperangular part is
\begin{eqnarray}
\Lambda^2&=&  -\frac{\partial^2}{\partial \alpha_i^2} - 
2\cot(2\alpha_i)\frac{\partial}{\partial \alpha_i}  \nonumber \\ \label{lambda} 
   &-&  \frac{1}{\sin^2\alpha_i}\frac{\partial^2}{\partial \Theta_{xi}^2} - 
\frac{1}{\cos^2\alpha_i}\frac{\partial^2}{\partial \Theta_{yi}^2}. 
\end{eqnarray}
The eigenfunctions corresponding to the dependence on $\Theta_{xi}$
and $\Theta_{yi}$ are $\exp(im_{xi} \Theta_{xi} +
im_{yi}\Theta_{yi})$, where $m_{xi}$ and $m_{yi}$ are integers.  The
sum, $M = m_{xi} + m_{yi}$, is a conserved quantum number which labels
the solutions by $M=0,\pm 1,\pm 2,...$.

The total wave function, $\Psi_M$, is expanded in a complete set of
hyperangular functions, $\Phi_{nM}(\rho,\Omega)$, for each $\rho$
chosen as eigenfunctions of the hyperangular part of the Hamiltonian,
$H = T + g\sum_{i=1}^3 V(\sqrt2 |{\bf x_i}|)$. Thus,
\begin{eqnarray}
 \Psi_M = \frac{1}{\rho^{3/2}}\sum_{n=0}^{\infty}
 f_{nM}(\rho)\Phi_{nM}(\rho,\Omega) \;, \label{totalwave}  \\ 
 \left(\Lambda^2+g\frac{2m\rho^2}{\hbar^2} \sum_{i=1}^{3} 
 V(\sqrt2 |{\bf x_i}|)\right)\Phi_{nM} = \lambda_{nM}(\rho)\Phi_{nM} , 
  \label{angular}
\end{eqnarray}
where $V$ is the two-body potential, and the functions $f_{nM}(\rho)$
satisfy a system of coupled equations.

We omit the index $n$ and search only for the angular solution,
$\Phi_M$, of lowest energy for a given $M$.  We use the Faddeev
decomposition and expansion on the eigenfunctions related to
$\Theta_{xi}$ and $\Theta_{yi}$, that is
\begin{eqnarray}\label{angfunc}
 \Phi_M  = \frac{1}{2\pi} \sum_{m_{xi},i} 
 \phi_{Mm_{xi}}(\rho,\alpha_i)
 e^{-im_{xi}\Theta_{xi}  - i(M - m_{xi})\Theta_{yi}},
\end{eqnarray}
where $i=1,2,3$ and the index $m_{xi}$ only assumes odd values to
ensure antisymmetry.

\begin{figure}
\includegraphics[scale=0.95]{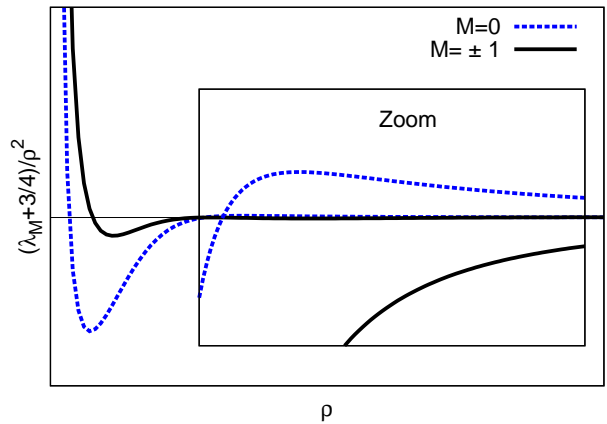}
\caption{Three-body potentials for ground state and first excited state channels.
Schematic illustration of the lowest adiabatic potentials for $M=0$ and $M=1$ at
short and long distance. The inset shows a zoom of the behavior at large distance to illustrate
the barrier in the $M=0$ channel. Note that only the vertical axis has been amplified in the inset.
The inset starts roughly at the distance $\rho\sim b$. For the potentials considered here
the attractive pocket in the $M=0$ channel is considerably larger than in the $M=1$
channel. We therefore have stronger binding in the $M=0$ channel.}
\label{fig-ad-pot}
\end{figure}

\subsection{Lowest order large-distance solution}
We first solve to lowest order without any non-adiabatic and coupling
terms, that is we find $f_{M}(\rho)$ in the so-called exact adiabatic
approximation \cite{coelho91}, from
\begin{eqnarray}
\left(-\frac{\partial^2}{\partial \rho^2} +\frac{\lambda_M+3/4}{\rho^2} 
- \frac{2mE_3}{\hbar^2}\right)f_M(\rho)  = 0 \;.
\label{eq-hyper}
\end{eqnarray}
Using the lowest adiabatic potential produced by $\lambda_M$ we then
get a lower bound on the energy \cite{coelho91}.  The behavior of
$\lambda_M$ for large $\rho$ is decisive for sufficiently weakly bound
states provided the centrifugal barrier is small.  

We find $\lambda_M$ by solving Eq.~\eqref{angular} with the structure
of $\Phi_M$ in Eq.~\eqref{angfunc}.  In general, the contributing
configurations in angular space arise from the smallest
$|m_{x}|$-values, since the centrifugal barrier in $\alpha$-space is
decided by the $1/(\sin^2\alpha)$-term in Eq.~\eqref{lambda}.  For our
cases of $M=0$ and $M=1$ these values are $(m_{x},m_{y}) =
(1,-1),(-1,1)$ and $(1,0),(-1,2)$, respectively.  Detailed
calculations of $\lambda_M$ are described in Methods.

With these $\lambda_M$ we obtain differential equations for
$f_M(\rho)$ from Eq.~\eqref{eq-hyper}, that is
\begin{eqnarray}
\left(-\frac{\partial^2}{\partial \rho^2} + \frac{3}{4 \rho^2} -\frac{16}{3\rho^2\ln(\rho/R_0)}
-\frac{2mE_3}{\hbar^2}\right)f_0(\rho)=0, \label{eq-fin-rho1} 
\\
\left(-\frac{\partial^2}{\partial \rho^2}-\frac{1}{4 \rho^2}-
\frac{16}{9 \rho^2 \ln^2(\rho/R_0)}-\frac{2mE_3}{\hbar^2}\right)f_1(\rho)=0,
\label{eq-fin-rho}
\end{eqnarray}
which are valid for strengths close to $g_{2}^{cr}$ where $a\to\infty$.

For $M=0$, the large-distance behavior is a repulsive centrifugal
barrier corresponding to two close-lying particles and the third far
away in a state of relative angular momentum $1$.  This barrier
excludes infinitely many bound states.
For $M=\pm 1$ the large-distance behavior is attractive with a leading
term arising from two particles close to each other and the third far
away with relative angular momentum $0$.  For $E_3=0$ an analytical
solution can be found as in Ref.~\cite{chadan} and easily proved by
insertion into Eq.~\eqref{eq-fin-rho}, that is
\begin{equation}
f_1(E_3=0,\rho)=\sqrt{\rho\ln{\rho}}\cos(s \ln(\ln(\rho/R_0))+\delta),
\end{equation} 
where $s=\sqrt{16/9-1/4}$ and the value of $\delta$ is related to the
short-distance boundary condition.  This is an oscillating function of
$\ln (\ln (\rho/R_0))$ with an infinite number of nodes.  Consequently
Eq.~\eqref{eq-fin-rho} has an infinite number of bound state solutions
with $E_3<0$.  

Each of these bound states of energy $E_3^{(n)}$ falls of
exponentially when $\rho^2$ increases above $\rho_n^2 =
\hbar^2/(2m|E_3^{(n)}|)$.  However, for $\rho < \rho_n$ all solutions
resemble $f_1(E_3=0,\rho)$, which therefore provides the estimate
$-E_3^{(n)} \propto \hbar^2/(2m \rho_n^2)$ where $\rho_n$ obey $s \ln
(\ln \rho_n/R_0) + \delta = \pi(n+1/2)$.  Thus, $-E_3^{(n)}\sim
\exp\left(-2e^{\pi n/s}\right)$ and we obtain a double exponential
scaling.  The simultaneous contributions from components,
$(m_{x},m_{y}) = (1,0),(-1,2)$, are crucial for this conclusion.

\subsection{Higher order effects}  
The derivation of $\lambda_M$ from
solving the angular equation, Eq.~\eqref{angular}, was very recently
modified in Ref.~\cite{gao2014} by including, in the angular wave function 
at small angles $\alpha$, the next order (faster
vanishing) term in $\rho$.  Surprisingly, this leads to an
additional slower vanishing, non-universal term,
$-Y/(\rho^2\ln(\rho/R_0))$, which should be added in both radial
equations Eqs.~\eqref{eq-fin-rho1} and \eqref{eq-fin-rho}.  For
details see Methods.

This unusual behavior suggests that other higher order terms also
should be investigated. First, we emphasize that only the angular
configurations with $|m_{x}| \le 1$ can contribute to a large-distance
long-range behavior supporting infinitely many bound states.  Second,
non-adiabatic coupling terms in the radial equations are expected to
vanish faster than the terms included so far.
Here the diagonal coupling term~\cite{nielsen2001}, $Q=\langle
\Phi_1|\frac{\partial^2}{\partial \rho^2}
|\Phi_1\rangle_{\{\Theta_x,\Theta_y,\alpha\}}$, is
of very special significance for two reasons.  First, it is
expected to be the slowest vanishing for large $\rho$, and second
omission or inclusion of this coupling in the lowest adiabatic
potential produces, respectively a lower or upper bound for the exact
binding energy \cite{coelho91}.  The leading order result for $Q$ is derived in
Methods to cancel exactly the non-universal term found in Ref.~\cite{gao2014}. 
As a byproduct $Y>0$ is simultaneously demonstrated.

Conclusions from these delicate considerations are first that the
$M=0$ channel essentially is unaffected, since this very small
additional term cannot change the large-distance main contribution, $3/(4\rho^2)$, of
the repulsive centrifugal barrier in Eq.~\eqref{eq-fin-rho1}.  Second,
it cannot change the short-distance attraction which is crucial for
borromean structures.  Third conclusion is that inclusion of these
terms are very important for the $M=1$ channel, since it may change the
large-distance behavior crucial for possible (super) Efimov states.

The necessary rigorous mathematical derivations and accurate
numerical investigations for $M=1$ are very involved and beyond the
scope of this paper.  However, if for instance, the next order of
$Q$ is equal to $1/(4\rho^2\ln^2(\rho/R_0))$ and the sum of all other
couplings vanish faster, the prediction in Ref.~\cite{superefimov}
would be confirmed.  We emphasize that this order in $Q$ is crucial
for the Efimov effect as it provides both upper and lower bounds on
the exact energy.

\begin{figure}
\includegraphics[scale=0.31]{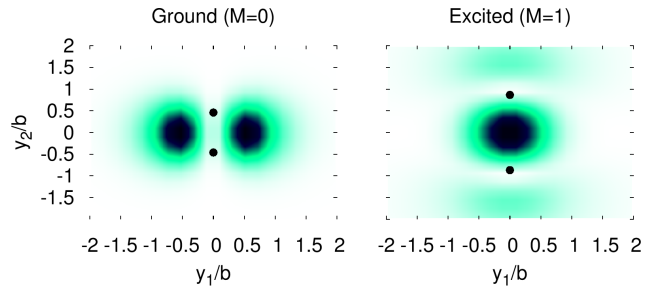}
\caption{Three fermion wave functions in ground and excited states.
	The ground (left) and excited (right) state probability
  distributions of the third fermion for $V_1$ at the two-body
  threshold as function of $\mathbf{y}$ for a fixed value of
  $\mathbf{x}= (0,\sqrt{ \langle x^2 \rangle})$ corresponding to the
  first two particles fixed at positions $(y_1,y_2) = (0,\pm \sqrt{
    \langle x^2 \rangle})/2$ indicated by the black dots. The
  probabilities increases from light (white) to dark (black) colors.
  The unit of length is $b$. }
\label{figground}
\end{figure}

\subsection{Borromean binding}
The ambiguity in the $M=1$ channel is present but not as important for the $M=0$
channel, where the short-distance behavior is decisive.  The effective
potential in Eq.~\eqref{eq-hyper} obtained from Eq.~\eqref{angular} is
for $\rho=0$ simply the eigenvalue, $K(K+2)$ of the kinetic energy
operator $\Lambda^2$, see Ref.~\cite{nielsen2001}. The Pauli principle 
selects $K=2,3$ for $M=0,1$, respectively. The potentials for small
$\rho$ are then $35/4\rho^2$ and $63/4\rho^2$.  Comparing the
potentials at short and large distances we conclude that the effective
potentials have to cross each other with more attractions in the $M=0$
channel at short distances. This behavior is plotted schematically in
Fig.~\ref{fig-ad-pot} which shows that the three-body ground state has
$M=0$.

Let us now consider strengths, $g$, such that $g^{cr}_3 \leq g \leq
g^{cr}_2$, where $g^{cr}_3$ is defined as the limit for binding three
fermions.  Since the wave function is fully antisymmetric we can use
the result in Ref.~\cite{richard1994} to obtain a lower bound for the
ground state energy, $E_3(m,g)\geq 2E_2(m,3g/2)$.  It then immediately
follows that $2/3 \leq g_3^{cr}/g_2^{cr} \leq 1$.  As for three bosons
in 3D, the lower limit is reached for deep and narrow two-body
potentials vanishing at large distance.

To investigate three-body systems numerically we 
apply the stochastic variational technique to the three-body
Schr{\"o}dinger equation using the basis elements
\begin{eqnarray}
\psi_i &=& (1-P_{12})(1-P_{13})(1-P_{23})G({\bf x},{\bf y}) \; , \\ \nonumber 
G &=& e^{-a_1(\mathbf{x}-\mathbf{s_1})^2-2a_2(\mathbf{x}-
\mathbf{s_1})(\mathbf{y}-\mathbf{s_2})-a_3(\mathbf{y}-\mathbf{s_2})^2}\; ,
\end{eqnarray}
where $P_{ij}$ is the operator that exchanges particle coordinates $i$
and $j$, and $a_k,\mathbf{s_k}$ are non-linear variational parameters
that are found stochastically~\cite{volthesis}.  In Table~\ref{tab} we
present the three-body energies, $E_3({g_2^{cr}})$, for the ground and
doubly degenerate first excited states.  Borromean binding is then
allowed for strengths smaller than $g_2^{cr}$ with corresponding
smaller three-body binding energies.  Away from $g_2^{cr}$ the Efimov condition
of infinitely large scattering length is violated.  The ground state
has $M=0$ for all the qualitatively different two-body potentials, and
energies and sizes are of the same order as the potential depths
and ranges.  The first excited states have much smaller binding
energy, much larger extension, and quantum number $M=\pm 1$.  They are
the first in the possible sequence of super-Efimov states.

The geometric structure of these states can be seen in the probability
density, $P=|\Phi(\mathbf{x},\mathbf{y})|^2$, shown in
Fig.~\ref{figground} for potential $V_1$.  We choose the
$\mathbf{x}$-coordinate to be $\mathbf{x}=(0,\sqrt{ \langle x^2
  \rangle})$ with a particle distance equal to the root-mean-square
value.  The structures in Fig.~\ref{figground} persist for the other
potentials and for different values of $|x|$.  The resulting
triangular and linear chain structures resemble the ground and the
celebrated first excited $0^+$ states for the $\alpha$-cluster
structure of the $^{12}$C-nucleus \cite{jen04}.  

By decreasing $g$ from $g_2^{cr}$ toward $g_3^{cr}$ we move into the
borromean window and away from infinite scattering length.  This quickly
removes all possible $M=\pm 1$ excited bound states.  However, the
$M=0$ ground state remains until the value $g=g_3^{cr}$ which is
calculated and given in Tab.~\ref{tab}.  Energies and sizes for
$g=g_{2}^{cr}$ given as mean-square radii, $\langle \rho^2\rangle$,
illustrate both stability and fragility of ground and excited states.
Ratios of their binding energies differ consistently with one order of
magnitude, and excited state radii are much larger than for the ground
state.  The effect of the outer barrier in $V_2$ is clearly seen
through larger binding and smaller radii.

\section{Discussion}
Using a hyperspherical formalism in 2D we have found that three
identical fermions support two very different types of bound states
corresponding to angular momentum $0$ and $1$.  We find a doubly
degenerate excited state of angular momentum one for two-body
potentials very close to the threshold for binding.  These states may
be the lowest in an infinite sequence of super Efimov states as
predicted in Ref.~\cite{superefimov}.  Our findings differ from
Ref.~\cite{gao2014} where a non-universal term appears in the effective
three-body potential. However, we show that this term is
cancelled precisely by inclusion of the leading order of the diagonal coupling.
In any case, the possible states are fragile, extremely large and
weakly bound, and observation of the related possible super Efimov
sequence would be correspondingly difficult.

A novel discovery is the existence of a borromean window defined as an
interval in finite two-body strengths where no two-body subsystem is
bound.  Within this window the well bound zero angular momentum ground
state is found to be located inside a barrier.  This is in sharp
contrast to the system of three bosons \cite{volbor}.  
Numerical calculations for three spin-polarized systems are carried
out for the first time in two dimensions using the coordinate space
formalism with finite-range potentials.  The results confirm the
analytically discussed properties for both ground and excited states.
These calculations can be used as a reference point for further
numerical studies, that could be performed with atomic potentials
obtained from state-of-the-art quantum chemical calculations.

The ground state Borromean states are located behind an outer barrier
as we have demonstrated here.  This implies 
that they should be
observable as a peak in the atom loss rate when the state is located
in the window for borromean binding precisely as the bosonic Efimov
effect has been observed
\cite{kraemer2006,pollack2009,zaccanti2009,gross2009}.  An alternative
way to probe and populate the states would be to use RF spectroscopy
\cite{lompe2010,nakajima2011,machtey2012}.  
It is imperative to notice
that we do not need the presence of Feshbach resonances in the angular
momentum one ($p$-wave) channel \cite{chin2010}. The method we have
used to derive the results above assumes only that there is a 
short-range two-body potential between the fermions that can be tuned
in some way so that one can approach the threshold for the existance of
a two-body bound state between two identical fermions. A 
$p$-wave Feshbach resonance is one option, but it could also be 
achieved by using for instance a long-range interaction such as 
for instance a dipole-dipole force.

\paragraph*{Acknowledgements}
Discussions with D.-W. Wang, J. Levinsen, S. Moroz, Y. Nishida and Z. Yu 
are gratefully acknowledged. This work was funded by the 
Danish Council for Independent Research DFF Natural Sciences and the 
DFF Sapere Aude program.

\appendix
\section{Two-body binding in two dimensions}
Here we discuss the behavior of the two-body binding energy near the threshold.
At this point we also would like to refer to some early works~\cite{gibson1986} using Jost function
formalism. For our derivations we use~\cite{vol11}
that the discrete spectrum of bound
states with $k=i|k|$ ($k^2=mE_2/\hbar^2$) corresponds to the solutions of the equation
\begin{eqnarray}  \label{jost-func}
 1+g k\frac{m}{\hbar^2}\frac{i\pi}{4}\int_0^\infty
\mathrm{d}r r\phi(k,r)V(r)H_1^{(1)}(k r) = 0\;,
\end{eqnarray}
where $H_1^{(1)}$ is the first Hankel function of order one~\cite{abram64}.  We focus on the
ground state in the weak binding limit, $k\to 0$, and write eq.~\eqref{jost-func} in the form
\begin{eqnarray}
&1+kg\dfrac{m}{\hbar^2}\dfrac{i\pi}{4}\int_0^\infty \mathrm{d}r r\phi(0,r)V(r)
 \bigg(\dfrac{i}{\pi}kr\big(\ln(kr/2)+\gamma \nonumber \\ &-1/2   -i\pi/2\big)-
\frac{2i}{\pi kr}\bigg) +o(k^2\ln kR_0)=0,
 \label{jost-small}
\end{eqnarray}
where $\gamma=0.577...$ is the Euler-Mascheroni constant
and where we use the regular zero energy solution from Eq.~(\ref{funct-small}).
Eq.~(\ref{jost-small}) with $k=0$ defines $g_2^{cr}$ as
the smallest solution of the equation:
\begin{equation}
1+g^{cr}_2\frac{m}{2\hbar^2}\int_0^\infty \mathrm{d}r \phi(r)V(r)=0 \;,
 \label{cond-gen}
\end{equation}
which from Eq.~(\ref{aeq}) is seen to correspond to the solution of
infinite scattering length, $a \to \infty$.  If $g$ is slightly larger
than $g_2^{cr}$ then the ground state is bound and the binding energy
satisfies the equation
\begin{equation}
AE_2\ln(|E_2|B)+g_2^{cr}-g\simeq0\; ,
\end{equation}
 where ($A,B,g_2^{cr}$) are positive constants. This dependency was
used to extract an accurate value of $g_2^{cr}$ by
fitting the numerically calculated $E_2$ as function of $g$.

\section{Three-body problem in the hyperspherical formalism}
Here we present the derivations that lead to the Eqs.(\ref{eq-fin-rho1}) and
(\ref{eq-fin-rho}). We show the approach to solve Eq.(\ref{angular})
with the wave function in Eq.(\ref{angfunc}).
The set $\phi_{M m_x}(\rho,\alpha)$, where for simplicity we use $i=1$ and
omit the related index,
solves the following system of integro-differential equations~\cite{nielsen2001}
\begin{align}
&\left(-\frac{\partial^2}{\partial \alpha^2} - 2\cot(2\alpha)
\frac{\partial}{\partial \alpha} + \frac{{m_x}^2}{\sin^2\alpha}+
\frac{m_y^2}{\cos^2\alpha}- \lambda_M\right) \phi_{M m_{x}}
\nonumber \\&= -g\frac{2m\rho^2}{\hbar^2}V(\sqrt{2}\rho\sin\alpha)\left(\phi_{M m_x}+\sum_{m_x'}
R_{M m_xm_x'}\right), \label{coupled_eq}
\end{align}
where $m_y=M-m_x$ and
the coupling term $R_{M m_xm_x'}$ is defined as
\begin{align}
R_{M m_xm_x'}&=\frac{1}{4\pi^2} \sum_{j\neq 1}\int \mathrm{d}\Theta_{x}
\mathrm{d}\Theta_{y} e^{im_x\Theta_{x}}e^{i(M-m_x)\Theta_{y}} 
\nonumber \\ &\times \phi_{M m_x'}(\rho,\alpha_j)e^{-im_x'\Theta_{xj}}e^{-i(M-m_x')\Theta_{yj}}.
\end{align}

{\it Non-interacting case.}
Let us first consider the situation when $V=0$, which also defines the
extreme short-distance behavior of $\lambda_M$ for potentials diverging slower
than $1/r^2$ at zero. In this situation the system of equations (\ref{coupled_eq})
decouples
\begin{align}
\bigg(-\frac{\partial^2}{\partial \alpha^2} - 2\cot(2\alpha)\frac{\partial}
{\partial \alpha} + \frac{m_x^2}{\sin^2\alpha}  +   \frac{m_y^2}{\cos^2\alpha}-\lambda_M\bigg)\phi^{(0)}_{M m_x}=0 \;,
\end{align}
where the superscript $0$ tells us that we work with free particles.
The solutions that are regular at $\alpha=0, \pi/2$ are~\cite{nielsen2001},
\begin{equation}
\phi^{(0)}_{M m_x} = N^{(0)}_{M m_x}\sin^{|m_x|}(\alpha)\cos^{|m_y|}(\alpha)
P_n^{(|m_x|,|m_y|)}(\cos(2\alpha)),
\end{equation}
where $P_n^{(|m_x|,|m_y|)}$ is the Jacobi polynomial~\cite{abram64}, $N^{(0)}_{Mm_x}$ is the
normalization constant, and $n$ is a non-negative integer related to
the eigenvalue $\lambda_M$ by $\lambda_M = K(K+2),\;  K = 2n +|m_x|+|M-m_x|$.
This result can also be used to determine the behavior
of $\lambda_M$ near $\rho=0$, since the right-hand side of Eq.(\ref{coupled_eq})
contains the factor $\rho^2$. From this we learn that for $M=1$ the lowest centrifugal
barrier is determined by $m_x=1, \;  n=1, \;  K=3$ (since the antisymmetric wave function with $m_x=1$
and $n=0$ is zero). For $M=0$ the lowest barrier arises
from $m_x=1, \; n=0, \;  K=2$.

{\it Interacting case.}  We aim to find the large-distance behavior of
$\lambda_M$.  We introduce an angle
$\alpha_0=R_0/(\sqrt{2}\rho)\rightarrow 0$, such that for
$\alpha>\alpha_0$ Eq.(\ref{coupled_eq}) is the non-interacting case
(right-hand side is zero) with the regular boundary condition at
$\alpha=\pi/2$ and solved by
\begin{align}
\phi_{M m_x}(\rho,\alpha)=N_{M m_x}(\rho)\sin^{|m_x|}(\alpha)\cos^{|M-m_x|}
(\alpha) \nonumber \\ \times P_{\nu_{M m_x}}^{(|M-m_x|,|m_x|)}(-\cos(2\alpha)),
\end{align}
where $\nu_{M m_x}(\rho)$ is given by $\lambda_{M}(\rho)=(2\nu_{M
  m_x}(\rho) + |m_x|+|M-m_x|)(2\nu_{M m_x}(\rho)+|m_x|+|M-m_x|+2)$.
This wave function diverges for $\alpha\to0$, where the interacting
solution has to be used.  To obtain the solution we use that only the
components with $|m_x|=1$ are necessary to lowest order, since higher
partial waves are suppressed at small interparticle distances,
i.e. $\alpha<\alpha_0$, equivalent to large distances. 
More precisely, in the absence of additional resonances in the 
higher partial wave channels with $|m_x|>1$
one can show that $\nu_{Mm_x}(\rho)=n+O((b/\rho)^{2|m_x|})$
where $n$ is an integer (corresponding to a free solution) \cite{nielsen2001}.
Such terms will therefore vanish much faster than the $|m_x|=1$ terms
and can thus be neglected.

{\it Case with $M=1$}.  For $M=1$
we need to solve the following system of equations for small $\alpha$
\begin{align}
\label{system_eq}
&\bigg(-\frac{\partial^2}{\partial \alpha^2} - 2\cot(2\alpha)
\frac{\partial}{\partial \alpha} \nonumber + \frac{1}{\sin^2\alpha}
-\lambda_1(\rho) \bigg)
\phi_{11}(\alpha,\rho) = \nonumber \\  &-\frac{2gm\rho^2}{\hbar^2}
V(\sqrt{2}\rho\alpha) (\phi_{11}(\alpha,\rho)+R_{111}+R_{11-1}),
\end{align}
\begin{align}
&\bigg(-\frac{\partial^2}{\partial \alpha^2} - 2\cot(2\alpha)
\frac{\partial}{\partial \alpha} +  \frac{1}{\sin^2\alpha}+
\frac{4}{\cos^2\alpha}  -
\lambda_1 \bigg) \phi_{1 -1} = \nonumber   \\
&-\frac{2gm\rho^2}{\hbar^2} V (\sqrt{2}\rho\alpha)(\phi_{1-1}+R_{1-11}+R_{1-1-1})\;.
\label{system_eq1}
\end{align}
We note that $\nu_{1-1}=\nu_{11}-1$,
since they should yield the same $\lambda_1$, i.e. $
(2\nu_{11}+1)(2\nu_{11}+3)=(2\nu_{1-1}+3)(2\nu_{1-1}+5)$.
For simplicity from now on we will write $\nu_{11}=\nu$. Now we expand the wave functions
$\phi_{M m_x'}(\rho,\alpha_j)e^{-im_x'\Theta_{xj}}e^{-i(M-m_x')\Theta_{yj}}$
with $j=2,3$ in the vicinity of $\alpha=0$ which leads to the coupling terms
\begin{widetext}
\begin{equation}
\begin{aligned}
&R_{111} \simeq  \alpha r_{11}N_{11}; \qquad r_{11}= -{_2}F_1(-\nu,\nu+2,1,1/4)-
\frac{3\nu(\nu+2)}{4}{_2}F_1(-\nu+1,\nu+3,2,1/4)  ; \\
&R_{11-1} \simeq \alpha r_{1-1}N_{1-1} ;\; r_{1-1}=
\frac{\Gamma(\nu+2)}{\Gamma(\nu)}\left(\frac{3}{4}
{_2}F_1(-\nu+1,\nu+3,3,1/4)-\frac{(\nu-1)(\nu+3)}{32}
{_2}F_1(-\nu+2,\nu+4,4,1/4)\right)  ; \\
&R_{1-1-1} \simeq \alpha r_{-1-1}N_{1-1} ; \; r_{-1-1}=
\frac{\Gamma(\nu+2)}{\Gamma(\nu)}\left(-\frac{{_2}F_1(-\nu+1,\nu+3,3,1/4)}{8}
-\frac{(\nu-1)(\nu+3)}{32}{_2}F_1(-\nu+2,\nu+4,4,1/4)\right);\\
&R_{1-11} \simeq  \alpha r_{-11}N_{11}; \qquad r_{-11}=
-\frac{3\nu(\nu+2)}{4}{_2}F_1(-\nu+1,\nu+3,2,1/4); \\
\end{aligned}
\end{equation}
\end{widetext}
where $\Gamma(x)$ is the gamma function and ${_2}F_1$ is the ordinary
hypergeometric function~\cite{abram64} where all couplings
$R\sim\alpha$ as for $|m_x|=1$.

The homogeneous part of Eqs.(\ref{system_eq}) and (\ref{system_eq1})
for small $\alpha$ has the main contribution from the two-dimensional
two-body equation, Eq.(\ref{shr-rad-eq}), as seen by using the substitution
$r=\sqrt{2}\rho\alpha$.  One solution to the inhomogeneous part is
$\phi_{11}=- (R_{11}+R_{1-1})$ and
$\phi_{1-1}=-(R_{-11}+R_{-1-1})$. In this way we find the solutions
for $\alpha<\alpha_0$
\begin{align}
\label{small_ang}
\phi_{11}=C\left(\rho\alpha-\frac{a^2}{2\rho\alpha}\right)-N_{11}\alpha r_{11}-N_{1-1}\alpha r_{1-1},
\\
\phi_{1-1}=C_1\left(\rho\alpha-\frac{a^2}{2\rho\alpha}\right)-N_{11}\alpha r_{-11}-N_{1-1}\alpha r_{-1-1},
\label{small_ang1}
\end{align}
where $C(\rho), C_1(\rho), N_{11}(\rho)$ and $N_{1-1}(\rho)$ are
functions that up to normalization should be determined by matching
the solutions and their first derivatives for $\alpha>\alpha_0$ and
$\alpha<\alpha_0$ at $\alpha_0$.  This matching can be done only for
specific values, $\lambda_1$, that can be be obtained from the
following equation

\begin{align}
\bigg(P_{\nu}^{(0,1)}(-\cos(2\alpha_0))+\frac{\alpha_0}{2}
\frac{\partial P_{\nu}^{(0,1)}(-\cos(2\alpha_0))}
{\partial \alpha}+r_{11}\bigg)\times \nonumber \\
\frac{\bigg(P_{\nu-1}^{(2,1)}(-\cos(2\alpha_0))+\frac{\alpha_0}{2}
\frac{\partial P_{\nu-1}^{(2,1)}(-\cos(2\alpha_0))}{\partial \alpha}
+r_{-1-1}\bigg)}{r_{-11}r_{1-1}}=1\; ,
\label{eigen-eq}
\end{align}
where we took the limit  $a\rightarrow \infty$, since it defines the two-body threshold
for binding. To simplify this
equation we use the following
identities, see for example Ref.~\cite{nielsen2001},
\begin{align}
P_{\nu}^{(a,b)}(-x)=   \cos(\pi\nu)P_{\nu}^{(b,a)}
(x)-\sin(\pi\nu)Q_{\nu}^{(b,a)}(x),
\end{align}
\begin{align}
 Q_{\nu}^{(1,0)}&(\cos(2\alpha))\simeq\frac{\Gamma(\nu+2)}{\pi\Gamma(\nu+1)}
\bigg[-2\gamma+1-\psi_{\Gamma}(1+\nu)- \nonumber \\ &\psi_{\Gamma}(\nu+2)-
2\ln(\alpha)\bigg]+\frac{\Gamma(\nu+1)}{\pi\Gamma(\nu+2)}\frac{1}{\alpha^2},
\\
 Q_{\nu-1}^{(1,2)}&(\cos(2\alpha))\simeq\frac{\Gamma(\nu+1)}{\pi\Gamma(\nu)}
\bigg[-2\gamma+1-\psi_{\Gamma}(\nu)- \nonumber \\ &\psi_{\Gamma}(\nu+2)-2\ln(\alpha)\bigg]
+\frac{\Gamma(\nu+2)}{\pi\Gamma(\nu+3)}\frac{1}{\alpha^2}\; ,
\end{align}
where  $\psi_{\Gamma}$ is the digamma function. These identities allow Eq.(\ref{eigen-eq})
to be rewritten
\begin{align}
\bigg(\frac{\Gamma(\nu+2)}{\Gamma(\nu+1)}\left(\cos(\pi\nu)+\frac{\sin(\pi\nu)}{\pi}
[\psi_{\Gamma}(1+\nu)+2\ln\alpha_0]\right)+r_{11}\bigg) \times \nonumber \\
\frac{\bigg(\frac{\Gamma(\nu+1)}{\Gamma(\nu)}\left(\cos(\pi\nu)+\frac{\sin(\pi\nu)}{\pi}
[\psi_{\Gamma}(\nu)+2\ln\alpha_0]\right)
+r_{-1-1}\bigg)}{r_{-11}r_{1-1}}=-1 \nonumber \; ,
\label{eigen-eq1}
\end{align}
where we neglect terms smaller than $|\ln\alpha_0|$ in the limit of $\alpha_0\to0$.
The smallest solution to this equation is $\nu=-1+\delta\nu$,
with $(\delta\nu)^2=-\frac{4}{9(\ln\alpha_0)^2}$.
This solution yields $\lambda_1=(2\nu+1)(2\nu+3)=-1-16/(9(\ln\alpha_0)^2)$, where
$\alpha_0\sim1/\rho$.
As we discuss in the main part of the paper this solution produces
the infinite tower of states with the double exponential scaling.

\vspace{1em}
\noindent {\it Validity of the derived large-distance behavior of $\lambda_1$ at $1/a=0$.}
Before applying this adiabatic potential
we need to discuss its validity.
To do so we first need to estimate what will be changed by adding higher
order terms in Eqs.(\ref{small_ang}) and (\ref{small_ang1}); and second
we need to calculate the lowest order coupling term~\cite{nielsen2001},
\begin{equation} \label{Diagcoupl}
Q=\langle \Phi_1|\frac{\partial^2}{\partial \rho^2}|\Phi_1 \rangle_{\Omega} \; ,
\end{equation}
where the averaging over all angles is taken for normalized angular
wave functions, $\langle \Phi_1|\Phi_1 \rangle_{\Omega}=1$.  First,
the higher order large-distance effects on the angular wave functions
are necessary, since their neglect is the only assumption made after
we decided to use only partial waves with $|m_x|=1$.  Second,
inclusion of the diagonal coupling yields an upper bound for the exact
binding energy, which together with the lower bound produced by
$\lambda_1$ establishes bounds for the exact three-body
binding energy.  It was very recently shown~\cite{gao2014} that the
higher order terms ($\sim1/\rho^2$) in the solution to the homogeneous
angular part of Eqs.(\ref{system_eq}) and (\ref{system_eq1})
contribute as $-Y/\ln(\rho/R_0)$ to $\lambda_1(\rho)$, where
\begin{equation}
Y=-1-\dfrac{\int_0^\infty\mathrm{d}x x^3 V(x) u(x)^2 }{\lim_{x\to\infty}(xu(x))^2}.
\end{equation}

This surprising effect can be understood since the boundary conditions
require that the constants in Eqs.(\ref{small_ang}) and
(\ref{small_ang1}) must satisfy $C(\rho)\sim \rho N_{11}(\rho)$.  This
means that a higher order term in Eqs.(\ref{system_eq}) and
(\ref{system_eq1}) may be of the same order as the couplings neglected
in Eqs.(\ref{small_ang}) and (\ref{small_ang1}).  In
Ref.~\cite{gao2014} it was pointed out that such an effect means
that two-body observables alone is not sufficient to reproduce the
correct asymptotic large-distance structure of $\lambda_1$.  Thus,
universal behavior is not guarantied.

To obtain an upper bound we estimate the diagonal coupling term, $Q$,
from Eq.(\ref{Diagcoupl}) as in Ref.~\cite{nielsen2001}.  For large $\rho$
we find up to terms of order $\sim
\left(\frac{1}{\rho^2\ln^2(\rho/R_0)}\right)$ the explicit expression:
\begin{widetext}
\begin{align}
&Q=C(\rho)\int_0^{\alpha_0} u(\sqrt{2}\alpha\rho)
\frac{\partial^2}{\partial \rho^2}
C(\rho)u(\sqrt{2}\alpha\rho) \sin(2\alpha)\mathrm{d}\alpha&\nonumber\\
&+N_{11}(\rho)\int_{\alpha_0}^{\pi/2} \sin(\alpha)
P_{\nu}^{(0,1)}(-\cos(2\alpha))
\frac{\partial^2}{\partial \rho^2}N_{11}(\rho)
 \sin(\alpha)
P_{\nu}^{(0,1)}(-\cos(2\alpha))\sin(2\alpha)\mathrm{d}\alpha,&
\end{align}
\end{widetext}
where $u(x)$ is the two-body wave function decreasing as $1/x$ at
infinity, the constants $C(\rho)=-\frac{\rho}{\sqrt{\ln(\rho)}}$ and
$N_{11}(\rho)=\frac{1}{\sqrt{2\ln(\rho)}}$ are determined to satisfy
normalization and boundary conditions at $\alpha=\alpha_0$.  After
straightforward but tedious calculation we obtain
$Q=-Y/(\rho^2\ln(\rho/R_0))$.  We know~\cite{nielsen2001} that $Q$ is
negative, and consequently $Y$ must be positive.  This can also easily
be confirmed directly by rewriting $Y$ as
\begin{equation}
Y=\dfrac{\int_0^\infty x\left[\dfrac{\partial u(x) x}{\partial x}\right]^2\mathrm{d}x}{\lim_{x\to\infty}(xu(x))^2}.
\end{equation}
This means that the derived adiabatic potential necessarily should be
supplemented with both the $Q$ term and the term,
$-Y/(\rho^2\ln(\rho/R_0))$, obtained by inclusion of next-to-leading
order terms in the solution to the homogeneous part of
Eqs.(\ref{system_eq}) and (\ref{system_eq1}).  It turns out that the
leading contribution in the $Q$ term cancels exactly the
$-Y/(\rho^2\ln(\rho/R_0))$ term.  Thus, to make a rigorous conclusion
about the allowed interval for the three-body binding energy, the
next-to-leading order in $Q$ term should be calculated.  This
investigation is, however, out of scope of the present paper but it
should be the focus in future more detailed and more elaborate work.

{\it Case with $M=0$}.  This case is not very different from the $M=1$
just discussed above.  We therefore only give the resulting
$\lambda_0$, and the couplings that are needed to solve the
corresponding equations.  The couplings for $\alpha\to 0$ take the
form
\begin{widetext}
\begin{align}
&R_{011} \simeq  \alpha r_{11}N_{01}; r_{11}= \frac{\Gamma(\nu+2)}{\Gamma(\nu+1)}\bigg(\frac{1}{2}{_2}F_1(-\nu,\nu+3,2,1/4)+\frac{3\nu(\nu+3)}{16}{_2}F_1(-\nu+1,\nu+4,3,1/4) \bigg) ;  \nonumber
\end{align}
\begin{align}
&R_{01-1} \simeq \alpha r_{1-1}N_{0-1} ;  r_{1-1}= \frac{\Gamma(\nu+2)}{\Gamma(\nu+1)}\bigg(-\frac{3}{2}{_2}F_1(-\nu,\nu+3,2,1/4)+\frac{3\nu(\nu+3)}{16}{_2}F_1(-\nu+1,\nu+4,3,1/4) \bigg) ; \nonumber \\
&R_{0-11} \simeq  \alpha r_{-11}N_{01}  ;  r_{-11}= \frac{\Gamma(\nu+2)}{\Gamma(\nu+1)}\bigg(-\frac{3}{2}{_2}F_1(-\nu,\nu+3,2,1/4)+\frac{3\nu(\nu+3)}{16}{_2}F_1(-\nu+1,\nu+4,3,1/4) \bigg);   \nonumber \\
&R_{0-1-1} \simeq \alpha r_{-1-1}N_{0-1} ;  r_{-1-1}= \frac{\Gamma(\nu+2)}{\Gamma(\nu+1)}\bigg(\frac{1}{2}{_2}F_1(-\nu,\nu+3,2,1/4)+\frac{3\nu(\nu+3)}{16}{_2}F_1(-\nu+1,\nu+4,3,1/4) \bigg);
\end{align}
\end{widetext}
where $\nu$ defines $\lambda_0$ as $\lambda_0=(2\nu+2)(2\nu+4)$.
In the same way as we obtained Eq.~(\ref{eigen-eq}) we derive
the following equation for $\nu$
\begin{align}
\bigg(P_{\nu}^{(1,1)}(-\cos(2\alpha_0))+\frac{\alpha_0}{2}
\frac{\partial P_{\nu}^{(1,1)}(-\cos(2\alpha_0))}{\partial \alpha}
+r_{11}\bigg)^2 =r_{-11}^2.
\label{eigen-eq2}
\end{align}
This equation has a particular solution $\nu=-1+\frac{4}{3
  \ln\alpha_0}$, which produces $\lambda_0=\frac{16}{3
  \ln\alpha_0}$. The corresponding adiabatic potential was presented
in the main part of this paper.  We also note that to determine the
large-distance behavior up to terms proportional to $1/\ln(\rho/R_0)$
we need an investigation similar to the one provided for $M=1$.
However, since the leading term in the adiabatic potential,
$3/(4\rho^2)$, only marginally allows universal weakly-bound states
the results of such calculations are not expected to be particularly
interesting.  In the $M=0$ channel the most important physics comes
from the non-universal adiabatic potential at short-distance.

\end{document}